\documentclass[twocolumn,5pt,prx,amsmath,superscriptaddress,nofootinbib,amssymb]{revtex4-1}
\usepackage{hyperref}
\usepackage{graphics}
\usepackage{amssymb}
\usepackage{amsmath}
\usepackage{upgreek}
\usepackage{graphicx}
\usepackage{times}
\usepackage{dcolumn}
\usepackage[usenames,dvipsnames]{color}
\usepackage{float}
\usepackage[utf8]{inputenc}

\usepackage[normalem]{ulem}
\usepackage{bm}
\usepackage{soul}
\usepackage{ragged2e}

\begin{document}
\renewcommand{\baselinestretch}{1}
\title{Engineering a Spin-Orbit Bandgap in Graphene-Tellurium Heterostructures}

\author{B. Mu\~{n}iz Cano}
\email{beatriz.muniz@imdea.org}
\affiliation{Instituto Madrile\~no de Estudios Avanzados, IMDEA Nanociencia, Calle Faraday 9, 28049, Madrid, Spain\looseness=-1}

\author{D. Pacilè}
\affiliation{Dipartimento di Fisica, Universit\`{a} della Calabria, Via P. Bucci, Cubo 30C, I-87036 Rende (CS), Italy\looseness=-1}

\author{M. G. Cuxart}
\affiliation{Instituto Madrile\~no de Estudios Avanzados, IMDEA Nanociencia, Calle Faraday 9, 28049, Madrid, Spain\looseness=-1}

\author{A. Amiri}
\affiliation{Instituto Madrile\~no de Estudios Avanzados, IMDEA Nanociencia, Calle Faraday 9, 28049, Madrid, Spain\looseness=-1}

\author{F. Calleja}
\affiliation{Instituto Madrile\~no de Estudios Avanzados, IMDEA Nanociencia, Calle Faraday 9, 28049, Madrid, Spain\looseness=-1}

\author{M. Pisarra}
\affiliation{Instituto Madrile\~no de Estudios Avanzados, IMDEA Nanociencia, Calle Faraday 9, 28049, Madrid, Spain\looseness=-1}
\affiliation{Dipartimento di Fisica, Universit\`{a} della Calabria, Via P. Bucci, Cubo 30C, I-87036 Rende (CS), Italy\looseness=-1}

\author{A. Sindona}
\affiliation{Dipartimento di Fisica, Universit\`{a} della Calabria, Via P. Bucci, Cubo 30C, I-87036 Rende (CS), Italy\looseness=-1}
\affiliation{INFN, sezione LNF, Gruppo collegato di Cosenza, Cubo 31C, I-87036 Rende (CS), Italy\looseness=-1}

\author{F. Mart\'{\i}n}
\affiliation{Instituto Madrile\~no de Estudios Avanzados, IMDEA Nanociencia, Calle Faraday 9, 28049, Madrid, Spain\looseness=-1}
\affiliation{Departamento de Qu\'{\i}mica, M\'{o}dulo 13, Universidad Aut\'{o}noma de Madrid, 28049 Madrid, Spain}

\author{E. Salagre}
\affiliation{Departamento de F\'isica de la Materia Condensada, Universidad Autónoma de Madrid (UAM), Campus de Cantoblanco, 28049, Madrid, Spain\looseness=-1}

\author{P. Segovia}
\affiliation{Departamento de F\'isica de la Materia Condensada, Universidad Autónoma de Madrid (UAM), Campus de Cantoblanco, 28049, Madrid, Spain\looseness=-1}
\affiliation{Condensed Matter Physics Center (IFIMAC), Universidad Autónoma de Madrid (UAM), Campus de Cantoblanco, 28049, Madrid, Spain\looseness=-1}
\affiliation{Instituto Universitario de Ciencia de Materiales ``Nicol\'as Cabrera", Universidad Autónoma de Madrid, Madrid, Spain\\
$\quad$
}

\author{E. G. Michel}
\affiliation{Departamento de F\'isica de la Materia Condensada, Universidad Autónoma de Madrid (UAM), Campus de Cantoblanco, 28049, Madrid, Spain\looseness=-1}
\affiliation{Condensed Matter Physics Center (IFIMAC), Universidad Autónoma de Madrid (UAM), Campus de Cantoblanco, 28049, Madrid, Spain\looseness=-1}
\affiliation{Instituto Universitario de Ciencia de Materiales ``Nicol\'as Cabrera", Universidad Autónoma de Madrid, Madrid, Spain\\
$\quad$
}

\author{A. L. Vázquez de Parga}
\affiliation{Instituto Madrile\~no de Estudios Avanzados, IMDEA Nanociencia, Calle Faraday 9, 28049, Madrid, Spain\looseness=-1}
\affiliation{Departamento de F\'isica de la Materia Condensada, Universidad Autónoma de Madrid (UAM), Campus de Cantoblanco, 28049, Madrid, Spain\looseness=-1}
\affiliation{Condensed Matter Physics Center (IFIMAC), Universidad Autónoma de Madrid (UAM), Campus de Cantoblanco, 28049, Madrid, Spain\looseness=-1}
\affiliation{Instituto Universitario de Ciencia de Materiales ``Nicol\'as Cabrera", Universidad Autónoma de Madrid, Madrid, Spain\\
$\quad$
}

\author{R. Miranda}
\affiliation{Instituto Madrile\~no de Estudios Avanzados, IMDEA Nanociencia, Calle Faraday 9, 28049, Madrid, Spain\looseness=-1}
\affiliation{Departamento de F\'isica de la Materia Condensada, Universidad Autónoma de Madrid (UAM), Campus de Cantoblanco, 28049, Madrid, Spain\looseness=-1}
\affiliation{Condensed Matter Physics Center (IFIMAC), Universidad Autónoma de Madrid (UAM), Campus de Cantoblanco, 28049, Madrid, Spain\looseness=-1}
\affiliation{Instituto Universitario de Ciencia de Materiales ``Nicol\'as Cabrera", Universidad Autónoma de Madrid, Madrid, Spain\\
$\quad$
}

\author{J. Camarero}
\affiliation{Instituto Madrile\~no de Estudios Avanzados, IMDEA Nanociencia, Calle Faraday 9, 28049, Madrid, Spain\looseness=-1}
\affiliation{Departamento de F\'isica de la Materia Condensada, Universidad Autónoma de Madrid (UAM), Campus de Cantoblanco, 28049, Madrid, Spain\looseness=-1}
\affiliation{Condensed Matter Physics Center (IFIMAC), Universidad Autónoma de Madrid (UAM), Campus de Cantoblanco, 28049, Madrid, Spain\looseness=-1}
\affiliation{Instituto Universitario de Ciencia de Materiales ``Nicol\'as Cabrera", Universidad Autónoma de Madrid, Madrid, Spain\\
$\quad$
}

\author{M. Garnica}
\email{manuela.garnica@imdea.org}
\affiliation{Instituto Madrile\~no de Estudios Avanzados, IMDEA Nanociencia, Calle Faraday 9, 28049, Madrid, Spain\looseness=-1}
\affiliation{Instituto Universitario de Ciencia de Materiales ``Nicol\'as Cabrera", Universidad Autónoma de Madrid, Madrid, Spain\\
$\quad$
}

\author{M. A. Valbuena}
\email{miguelangel.valbuena@imdea.org}
\affiliation{Instituto Madrile\~no de Estudios Avanzados, IMDEA Nanociencia, Calle Faraday 9, 28049, Madrid, Spain\looseness=-1}

\begin{abstract}
Intensive research has focused on harnessing the potential of graphene for electronic, optoelectronic, and spintronic devices by generating a bandgap at the Dirac point and enhancing the spin-orbit interaction in the graphene layer. Proximity to heavy \textit{p} elements is a promising approach; however, their interaction in graphene heterostructures has not been as intensively studied as that of ferromagnetic, noble, or heavy \textit{d} metals, neither as interlayers nor as substrates. In this study, the effective intercalation of Te atoms in a graphene on Ir(111) heterostructure is achieved. Combining techniques such as low energy electron diffraction and scanning tunneling microscopy, the structural evolution of the system as a function of the Te coverage is elucidated, uncovering up to two distinct phases. The presented angle-resolved photoemission spectroscopy analysis reveals the emergence of a bandgap of about 240 meV in the Dirac cone at room temperature, which preserves its characteristic linear dispersion. Furthermore, a pronounced n-doping effect induced by Te in the heterostructure is also observed, and remarkably the possibility of tuning the Dirac point energy towards the Fermi level by reducing the Te coverage while maintaining the open bandgap is demonstrated. Spin-resolved measurements unveil a non-planar chiral spin texture with significant splitting values for both in-plane and out-of-plane spin components. These experimental findings are consistent with the development of a quantum spin Hall phase, where a Te-enhanced intrinsic spin orbit coupling in graphene surpasses the Rashba one and promotes the opening of the spin-orbit bandgap.\\

\end{abstract}

\maketitle

\renewcommand{\baselinestretch}{0.5}
\centering
\section*{Introduction}   
\justifying

Since the discovery of the unique properties of graphene (Gr) such as its high electron and hole mobilities \cite{bae2010roll,ryu2014fast}, the long spin lifetime and long-distance spin propagation at room temperature (RT) \cite{avsar2019colloquium}, or its high electrical conductivity \cite{peres2010colloquium}, research into its properties has experienced a great increase. The Gr electronic structure is characterized by its $\uppi$ valence band, which is half filled and degenerate at the $\overline{\textrm{K}}_{\textrm{Gr}}$ point exactly at the Fermi level (E$_{\text{F}}$), the so-called Dirac point (DP). In the vicinity of the DP its band structure follows a linear relativistic-like dispersion described by the Dirac equation. In spite of these exceptional features, the application of Gr to electronic, spintronic or spinorbitronic devices  \cite{fiori2014electronics,han2014graphene,dlubak2012highly,ajejas2018unraveling} has been jeopardized by the fact that pristine Gr lacks a bandgap, difficulting the control of certain technologically relevant processes such as on-off switching operations in transistors \cite{chaves2020bandgap}.

Three main mechanisms have been explored to induce the desired bandgap in Gr, namely: i) the imposition of a superperiodicity on the Gr lattice, leading to confinement effects \cite{balog2010bandgap,sprinkle2010scalable,papagno2012large,son2006half,deParga2008periodically}; ii) the breaking of the sublattice symmetry and the subsequent generation of a lattice mismatch \cite{yavari2010tunable,zhou2007substrate,enderlein2010formation,krivenkov2021origin,warmuth2016band,bostwick2007symmetry}; and iii) the development of a sufficiently strong spin-orbit interaction (SOI) \cite{klimovskikh2017spin,otrokov2018evidence, calleja2015spatial,ma2011first}. 

In this regard, some technologically significant quantum states \cite{bader2010spintronics}, such as the quantum spin Hall (QSH) effect \cite{kane2005z}, require not only the opening of a gap, but also the sublattice symmetry not to be broken, thus preserving time-reversal symmetry (TRS). The QSH phase is a two dimensional (2D) topological insulator state, which has been predicted to be achievable in Gr \cite{kane2005quantum,hasan2010colloquium,qi2011topological}. In this context, the SOI must be taken into account for the generation of a bandgap. When the SOI is strong enough, the bulk bandgap of Gr is inverted, allowing for the realization of a topologically non-trivial state. If the strength of the intrinsic SOI is larger than the Rashba SOI, according to Kane and Mele's prediction, a bandgap appears at the Gr DP.  Nevertheless, the intrinsic spin-orbit splitting of pristine Gr is extremely weak and, unavoidably, overcome by the Rashba SOI \cite{rashba2009graphene,neto2009impurity,qiao2010quantum,sanchez2010chemical,boettger2007first}, resulting in a insignificantly small energy gap \cite{min2006intrinsic,yao2007spinorbitgap}.

Different routes have been pursued to enhance the SOI in Gr and, therefore, to promote the spin-orbit gap. Of significant relevance is the attempt to induce a strong hybridization with an intercalated heavy metal (HM) \cite{dedkov2015graphene,voloshina2014general}. Particularly, much research has been devoted to 5\textit{d} HM intercalation \cite{klimovskikh2015variation,vita2014understanding,zhang2012electrically, hu2012giant}. However, these kind of metals (such as Au or Ir) only increase the Rashba spin splitting of the electronic states \cite{shikin2013induced,marchenko2013spin,varykhalov2015tunable}. On one hand, one of the most successful cases for enhancing the SOI and generating the spin-orbit gap has been the intercalation of \textit{p} outer shell HMs \cite{brey2015spin}, as Pb on Gr/Pt(111) \cite{klimovskikh2017spin} or Gr/Ir(111) \cite{otrokov2018evidence}. In Gr/Pb/Ir(111) the extrinsic SOI continued to overcome the intrinsic one, and no gap opening was observed. In the case of Gr/Pb/Pt(111), a surpassing of the SOI was achieved, resulting in the opening of a spin-orbit gap. On the other hand, similar \textit{p} orbital HMs intercalation, as Bi on Gr/Ir(111) \cite{krivenkov2021origin}, do not promote any sizeable SOI effects in Gr and the observed gap opening at the DP has been related to the breaking of the sublattice symmetry, which prevents this system to be used as a platform for the implementation of the QSH effect. Thus, while 5\textit{d} HMs have been studied in much depth, the influence of the last \textit{p} filling shell is less known. Then, continuing the trend of the already studied Pb and Bi (as well as the also investigated Sn \cite{briggs2020atomically} and Sb \cite{lin2022vertical}) by following the band filling of the outer \textit{p} shell metals, the present work aims at investigating the electronic structure of Te-intercalated Gr/Ir(111) heterostructures. Moreover, this route provides very valuable information about the possible interactions between chalcogen atoms and Gr supporting substrates in the context of engineering novel 2D materials such as transition metal dichalcogenide (TMD) monolayers (ML) and other van der Waals (vdW) heterostructures with tailored properties \cite{gmitra2016trivial,roldan2014momentum,avsar2014spin, qiu2022resurrection}.

In the first part of this work, low energy electron diffraction (LEED) and scanning tunneling microscopy (STM) are used to demonstrate how the Te intercalation process can be successfully performed in Gr/Ir(111). Two long-range ordered structural phases are unveiled depending on the intercalated Te coverage. In the second part, the effects induced in the Gr electronic structure by the intercalated Te phases are also elucidated by means of angle-resolved photoemission spectroscopy (ARPES), spin-resolved ARPES (SR-ARPES) and density functional theory (DFT) calculations. Particularly significant is the emergence of a substantial bandgap at the DP at RT extending up to 240 meV and its tuning with the Te coverage. Interestingly, at low Te coverage the bandgap persists, and Gr becomes nearly charge neutral, in such a way that the E$_{\textrm{F}}$ is tuned into the gap. The SR-ARPES measurements evidence the intrinsic SOI origin of the induced gap, opening up new viable routes for the implementation of electronic or spintronic devices based on Gr/Te heterostructures.

\centering
\section*{Results and discussion}
\subsection*{Structural characterization} 
\justifying

\begin{figure*}
	\centering
\includegraphics[width=0.98\textwidth]{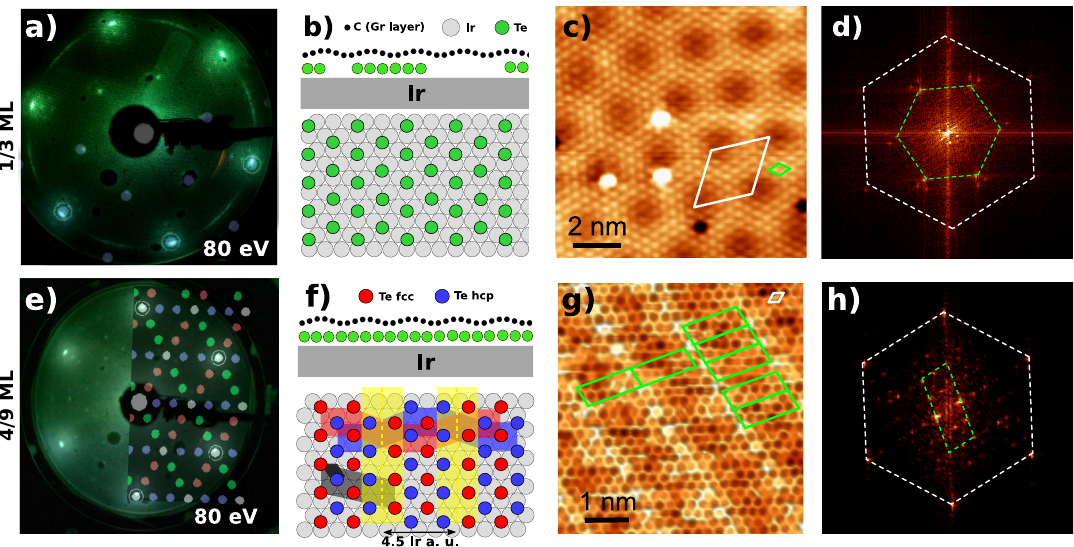}
	\caption{\textbf{LEED and STM characterization of the Te-intercalated Gr/Ir(111) heterostructures revealing two different structral phases depending of the amount of intercalated Te.} Top panels: 1/3 ML phase. (a) LEED pattern showing a pure ($\sqrt{3}$$\times$$\sqrt{3}$)R$30^{\circ}$ reconstruction (blue superimposed spots), described by the real space model in (b). (c) High resolved (HR) STM images (V$_\text{b}$=2 V, I$_\text{t}$=0.35 nA). The moiré pattern unit cell and the one created by the Te are depicted in white and green, respectively. (d) FFT image of (c) indicating a $(\sqrt{3}$$\times$$\sqrt{3})$R$30^{\circ}$ structure (green dashed hexagon, white one corresponds to Gr). Bottom panels: 4/9 ML phase. (e) LEED pattern and (f) associated real space model showing a reconstruction (whose three different domains are superimposed in blue, red and green on top of the LEED) consisting in a series of alternated periodic repetitions of a c(4x2) reconstruction in which the maximum number of Te atoms is accommodated. The unit cells are also depicted, following the color correspondence, along with the ones from Ir(111) and the Te reconstruction, in dark and light grey, respectively. Yellow rectangles indicate the areas which give rise to the 1D stripes in the STM experiments separated by 4.5 Ir atomic units (a.u.). (g) HR-STM images (V$_\text{b}$=1 mV, I$_\text{t}=$0.3 nA). The Gr lattice (white unit cell) can be seen superimposed on the stripe pattern created by the Te intercalation (green unit cell). Subtle 3D rendering has been applied for better visualization. (h) FFT pattern corresponding to (g), showing the same reconstruction as the one in the LEED in (e). LEED patterns have been acquired at E = 80 eV. In the real space models, grey circles represent Ir atoms and green ones stand for Te in (b), where the light green rhomboid depicts the unit cell of the structure. In (f), both red and blue circles represent Te in fcc and hcp positions with respect to the Ir surface, respectively (see SM, Section V).}
	\label{Fig1}
\end{figure*}

The samples were prepared in three steps, as described in more details in the Experimental Methods section. First, a Gr ML was grown onto an Ir(111) single crystal by chemical vapor deposition, whose LEED exhibiting the expected moiré is presented in Fig.\,S1(a) in the Supplementary Material (SM). Secondly, different amounts of Te were evaporated on Gr/Ir(111), promoting the formation of Te islands. As a result, the moiré spots are suppressed and several concentric circles indicating the multiple domains of the Te islands appear at the selected electron beam energy (80 eV, Fig. S1(b) in the SM). Finally, after annealing at 473 K the samples, the process culminates in the effective intercalation of Te between Gr and the supporting Ir(111) substrate. Depending on the Te coverage, two different phases are obtained. The LEED patterns associated to such phases are shown in Figs.\,\ref{Fig1}(a) and (e), respectively. Both LEED images show different spot distributions from those in Figs.\,S1(a) and (b) in the SM, indicating the complete Te intercalation and the formation of a superstructure of Te over Ir(111). Note that the LEED in  Fig.\,\ref{Fig1}(a) shows a reminiscence of the moiré pattern, which indicates a slightly reduced decoupling of Gr from the substrate. This can be  justified by the lower concentration of Te in the 1/3 ML phase compared to the 4/9 ML phase, where it is not observed. The matrices describing these reconstructions are given by:
\begin{equation}
    \notag G_{(1/3)} = 
\big(\begin{smallmatrix}
  2 & -1\\
  -1 & 2\
\end{smallmatrix}\big)
\quad \quad G_{(4/9)} = \big(\begin{smallmatrix}
  4 & 1\\
  -1 & 2\\
\end{smallmatrix}\big)
\end{equation}
 
whose subindexes are explained in the following. Also, models of the real space structures associated with such patterns are presented in Figs.\,\ref{Fig1}(b) and (f), correspondingly. Thus, two different regimes are identified, namely, a ($\sqrt{3}$$\times$$\sqrt{3}$)R$30^{\circ}$ phase (Figs.\,\ref{Fig1}(a) and (b)), and a series of alternated periodic repetitions of a c(4$\times$2) reconstruction in which the maximum number of Te atoms is accommodated (Figs.\,\ref{Fig1}(e) and (f)).
In a preliminary estimation of the density of Te atoms extracted from the real space structure, Te/Ir ratios of $1/3$ ML and of $4/9$ ML are found, which is the notation we will refer to from here onwards. In all reconstructions, Te atoms (in green in Fig.\,\ref{Fig1}(b) and in both red and blue in Fig.\,\ref{Fig1}(f), see figure caption for more details) have been placed at threefold on-hollow sites with respect to the Ir lattice since these are the most energetically favorable adsorption sites (see Section V in the SM). The same structural configuration has been predicted in similar systems \cite{bernardo2016intercalation,pisarra2018coverage,pisarra2018electronic}. 

A high resolution (HR) STM topographic image of the $1/3$ ML sample acquired in one of the Te-intercalated areas is shown in Fig.\,\ref{Fig1}(c). A characteristic large scale STM image is included in Fig.\,S2(a) in the SM, whose profile along the black line (displayed in the lower panel) reveals different flat areas with a step height associated to Ir atomic terraces (2.2 $\textup{\r{A}}$) and small patches with a height compatible with Te intercalation (1.6 $\textup{\r{A}}$), in the latter of which Fig.\,\ref{Fig1}(c) was acquired. Both Fig.\,\ref{Fig1}(c) and its corresponding fast Fourier transform (FFT) image (Fig.\,\ref{Fig1}(d)) reveal how the Te atoms are arranged in a ($\sqrt{3}$$\times$$\sqrt{3}$)R$30^{\circ}$ reconstruction, in agreement with the LEED in Fig.\,\ref{Fig1}(a). In addition, a closer inspection in one of these areas (see Fig.\,S3(a) in the SM) reveals intra-valley scattering around point defects promoted by the Te intercalation \cite{kuhne2017ultrafast,mao2012silicon,wang2021direct}.

\begin{figure*}
	\centering
        \includegraphics[width=0.98\textwidth]{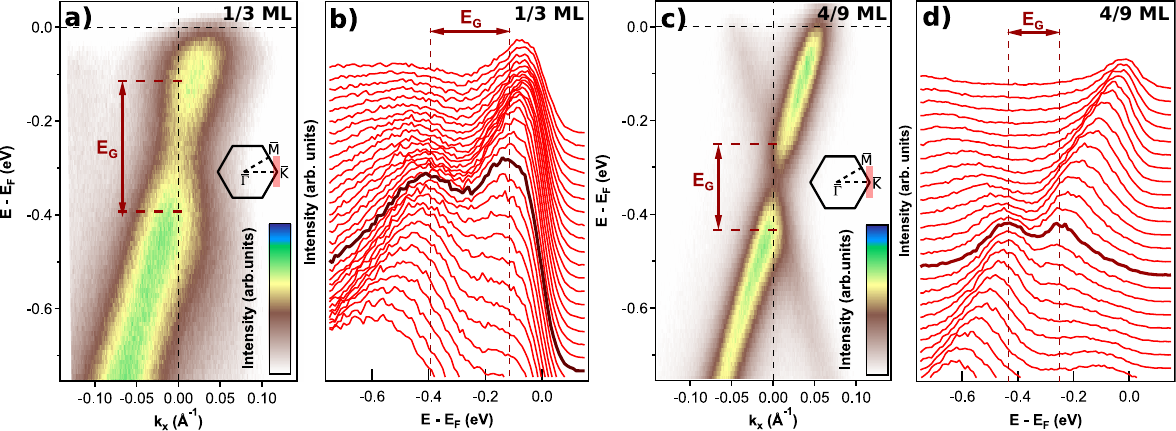}
	\caption{\textbf{ARPES study on the Te-coverage dependence of the Dirac cone in the Dirac point (DP) vicinity (h$\upnu$ = 21.2 eV and RT).} ARPES bandmaps close to the DP and measured at the $\overline{\text{K}}_{\textrm{Gr}}$ point along the $\overline{\Gamma \text{K}}$ perpendicular direction for the (a) $1/3$ ML and (c) $4/9$ ML Te samples. A sketch of the experimental geometry is included in (a) and (c). EDCs for the (b) $1/3$ ML and (d) $4/9$ ML Te samples  obtained from (a) and (c), respectively. Dark red EDCs correspond to the $\overline{\text{K}}_{\textrm{Gr}}$ point. An energy shift of the DP away from the E$_{\text{F}}$ as the Te coverage is increased is observed with values of, accordingly, (260 $\pm$ 80) meV and  (340 $\pm$ 30)  meV. The width of the bandgap E$_{\text{G}}$ amounts for (240 $\pm$ 80) meV and (180 $\pm$ 30) meV for the $1/3$ ML and $4/9$ ML Te phases, respectively.}
	\label{Fig2}
\end{figure*}

In the $4/9$ ML sample, most of the surface shows a set of quasi-one dimensional stripes oriented in three preferential directions (see Figs.\,S2(b) and S3(b) in the SM). A HR-STM image recorded on the terraces and its corresponding FFT pattern are shown in Figs.\,\ref{Fig1}(g) and (h), respectively. The absence of the moiré pattern and the fact that the Gr can be clearly resolved indicate that the Te is intercalated. The separation between the stripes is of $(1.3 \pm 0.1)$ nm $\approx$ 4.5 Ir atomic units (a.u.). This value is in good agreement with the real space model represented in Fig.\,\ref{Fig1}(f). The same consistency is found between the remaining STM findings, as compared to the conclusions previously drawn from the LEED characterization.

\centering
\subsection*{Electronic structure characterization} 
\justifying

A systematic ARPES study of the two Gr/Te/Ir(111) heterostructures performed with He I radiation was carried out to characterize the evolution of the band structure at RT. High-resolution ARPES spectra of the two representative samples ($1/3$ ML and $4/9$ ML) at the Dirac cones, just below the E$_{\textrm{F}}$, are displayed in Figs.\,\ref{Fig2}(a) and (c). The measurements were acquired at the $\overline{\text{K}}_{\textrm{Gr}}$ point along the perpendicular direction to $\overline{\Gamma \text{K}}$ (as schematically depicted in the insets). The overall band structures in an energy range spanning from the E$_{\textrm{F}}$ to the bottom of the Gr $\upsigma$ band are included in Fig.\,S4 in the SM, in which the band structure of an even lower Te coverage sample can also be found. Gr $\uppi$ bands are very sharp and intense, which indicates an optimal structural and interfacial quality. No signs of the Gr $\uppi$ band replicas arising from the moiré pattern can be appreciated (except from the bandmap corresponding to the lower coverage in Fig.\,S4(d), see the SM for details), something that, in conjunction with the subsequent absence of avoid-crossing mechanism bandgaps, indicates an effective Gr decoupling from the Ir(111) substrate due to the Te intercalation. Most importantly, the ARPES bandmaps in Figs.\,\ref{Fig2}(a) and (c) reveal the emergence of a bandgap at the DP in both of the studied phases.

To characterize such a bandgap opening and the induced n-type doping, the energy distribution curves (EDCs) are presented in Figs.\,\ref{Fig2}(b) and (d). Here, the opening of a bandgap at the DP, as well as its shifting towards higher binding energies (BE) with respect to Gr/Ir(111) are evident. Also, the quasi-free-standing character of the Gr $\uppi$ band is proved with an estimated group velocity, obtained from $\text{1/}\hbar \text{ }(\partial \text{E}/ \partial \text{k})$ \cite{miyamoto2012topological}, of (0.98 $\pm$ 0.04)$\cdot 10^{6}$ m/s, comparable to the reported values for Gr/Ir(111), of 0.95$\cdot 10^{6}$ m/s \cite{kralj2011graphene}, or free-standing Gr, of $10^{6}$ m/s. 

The width of the bandgap E$_{\text{G}}$ amounts for (240 $\pm$ 80) meV and (180 $\pm$ 80) meV for increasing Te coverage. Notably, the energies of the DP are (260 $\pm$ 80) meV and (340 $\pm$ 80) meV, respectively, for the same increasing Te coverage. The bandgap has been calculated as the difference between the two maxima associated to the $\uppi$ and $\uppi$* bands, and E$_{\textrm{DP}}$ as the middle energy between them. This demonstrates that a controllable n-type doping of the system can be achieved while, most importantly, the bandgap is preserved for both the studied phases up to RT.

Finally, in order to investigate whether the bandgap is tunable up to the E$_{\textrm{F}}$, an even lower Te coverage sample was studied after reducing the Te evaporation dose. This sample exhibits the same LEED pattern as in Fig.\,\ref{Fig1}(a). Figures S4 (c) and (f) in the SM demonstrate how this sample is almost charge neutral, with the DP close to the E$_{\textrm{F}}$. The bandgap can be better revealed by a small n-type doping through Na alkali deposition on top of Gr. As shown in Fig.\,S5 in the SM, the bandgap is preserved and its energy can be consequently tuned backwards by such an electronic charge transfer.

\begin{figure}
    \centering
    \includegraphics[width=0.48\textwidth]{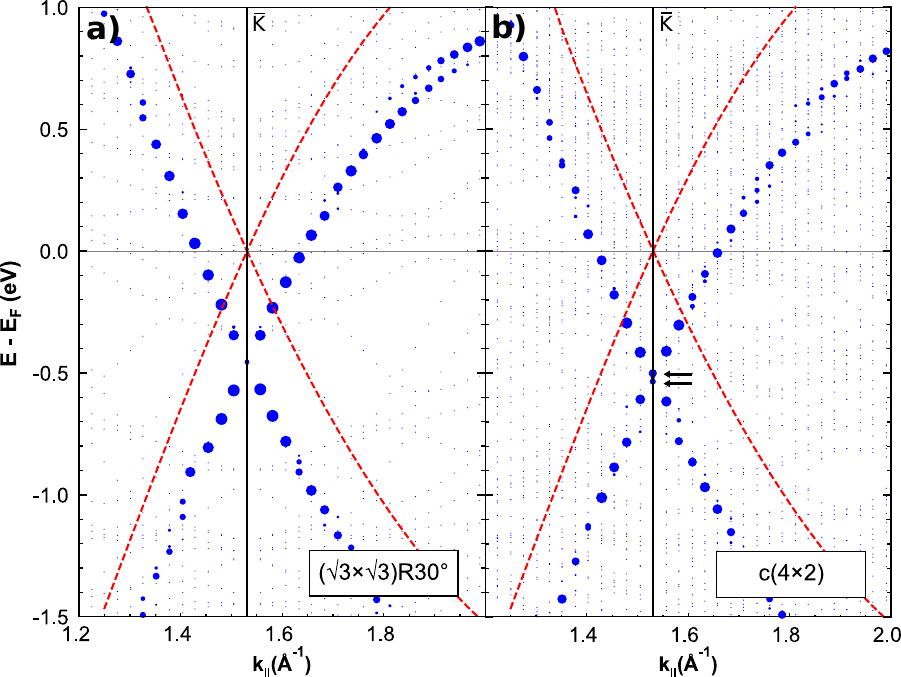}
	\caption{\textbf{DFT computed band structure near the Dirac point.} (a) Band-unfolded and Gr-projected band structure near the $\bar{\textrm{K}}_{\textrm{Gr}}$ point and the Fermi level for the $(\sqrt{3}\times \sqrt{3})$R$30^{\circ}$ geometry, corresponding to a $1/3$ ML Te coverage. A 6 meV (not visible) gap is found at the $\bar{\textrm{K}}_{\textrm{Gr}}$ point. (b) Same as (a) for the perfectly periodic c$(4\times2)$ geometry, corresponding to a $1/2$ ML Te coverage. A double point with high weight, which shows that a $\sim28$ meV gap (black arrows) is opened at the vertex of the Dirac cone, can be observed at the $\bar{\textrm{K}}_{\textrm{Gr}}$ point.}
 \label{Fig3}
\end{figure}

To further clarify the effect of the Te intercalation in the Gr band structure, a DFT adsorption study of Gr on several partially Te covered Ir surfaces was performed. In order to keep the size of the computations manageable, a stretched Gr layer was placed on two different Te covered Ir(111) surfaces so that the Gr lattice vectors match the Ir(111) lattice vectors, namely the $(\sqrt{3}\times \sqrt{3})$R$30^{\circ}$, characterized by a $1/3$ ML Te coverage; and the perfectly 2D periodic c$(4\times2)$ reconstruction, characterized by a $1/2$ ML Te coverage (see the Experimental Methods, and Section V in the SM for more details). In both cases, the structural optimizations evidence that the Gr-substrate interaction has a dispersive nature, with the C atoms stabilized at a $\sim3.5$\,\AA\, vertical distance from the Te atoms. As for the electronic properties, upon inspection of the band-unfolded and Gr projected band structure in Figs.\,S7 and S8 in the SM, the Gr band dispersion is found to be almost unaffected by the presence of the substrate which, however, induces a significant n-type doping. A closer look shows that Gr-substrate band hybridization is present in several places, as shown by the band splitting of the Gr derived bands, even though the linear dispersion of the $\uppi$ bands is preserved in the vicinity of the $\overline{\textrm{K}}_{\textrm{Gr}}$ point. In Fig.\,\ref{Fig3} a zoom-in close to the $\overline{\textrm{K}}_{\textrm{Gr}}$ point and the E$_{\textrm{F}}$ for the band structure of the two investigated intercalation regimes is reported. Here, the vertex of the Dirac cone is clearly downwards shifted by $\sim 470$ meV and $\sim 520$ meV for the $(\sqrt{3}\times \sqrt{3})$R$30^{\circ}$ and c$(4\times2)$ reconstructions, respectively, in fairly good agreement with the experimental results. It is also important to note that, in both cases, a small band gap is opened at the Dirac cone.  For the $(\sqrt{3}\times \sqrt{3})$R$30^{\circ}$ a very small $\sim6$ meV gap is found, not visible in Fig. \ref{Fig3}\,(a).  For the c$(4\times2)$ a $\sim28$ meV gap is spotted in Fig. \ref{Fig3}\,(b) Additional considerations on the bandgap opening from calculations with SOC can be found in Section V in the SM. 

\centering
\subsection*{Spin-resolved electronic characterization}
\justifying

\begin{figure*}
	\centering
        \includegraphics[width=0.98\textwidth]{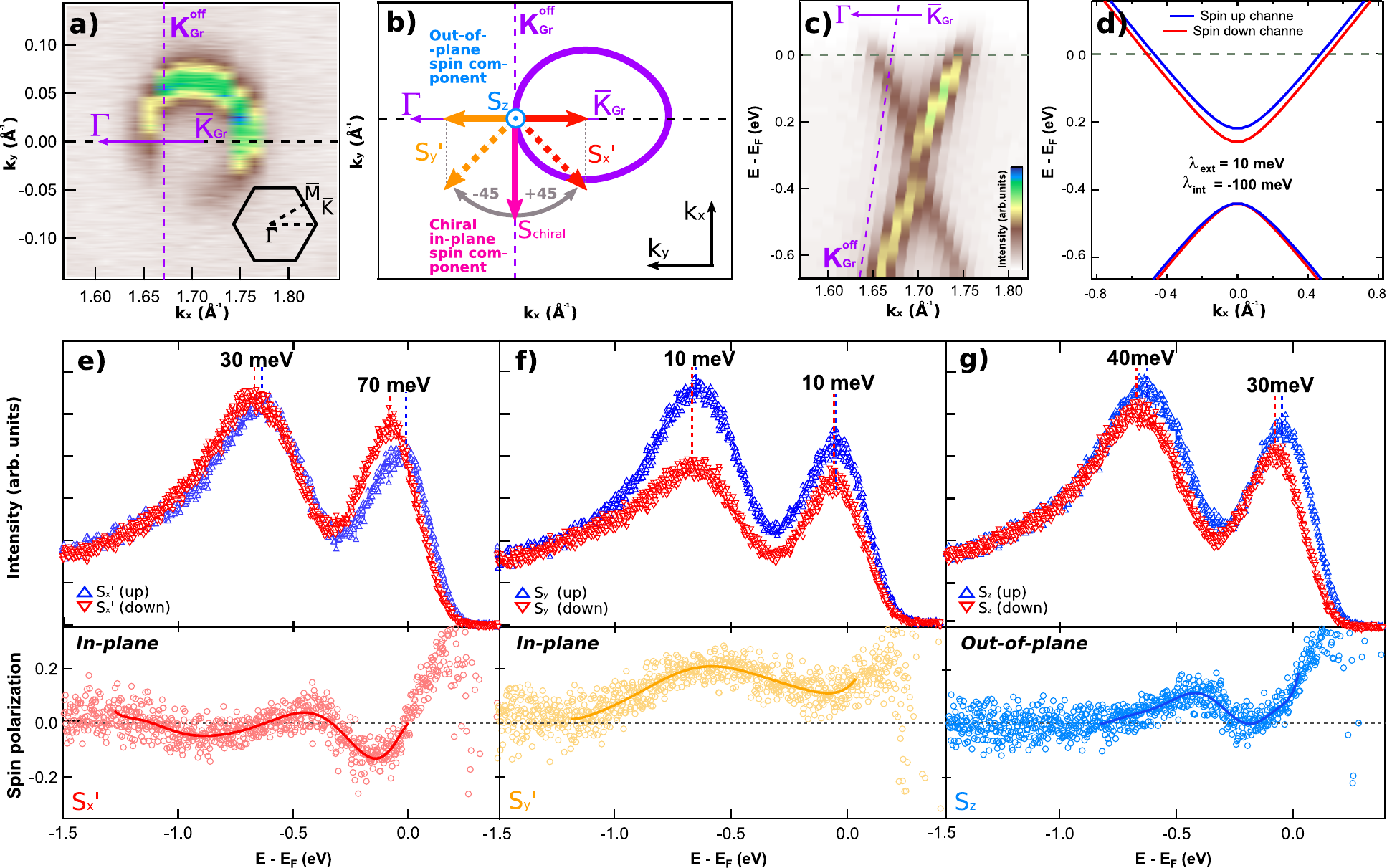}
	\caption{\textbf{ARPES and SR-ARPES study of the 4/9 ML heterostructure at RT: gap opening and non-planar chiral spin texture of the gapped graphene Dirac cone.} (a) Experimental FS around $\overline{\textrm{K}}$$_{\text{Gr}}$ point. The inset indicates the measured portion of the SBZ. The vertical purple dashed line corresponds to the momentum at which SR-ARPES measurements were recorded. The horizontal one indicates the direction for the ARPES bandmap shown in panel (c). (b) Schematical sketch of the experimental FS to illustrate the effect of the spin rotator lens. A net chiral in-plane spin component (fucsia arrow) is rotated $\pm$ 45$^\circ$. Thus, two orthogonal in-plane spin polarizations ($S_{x^\prime}$, red dashed arrow; and $S_{y^\prime}$, yellow dashed arrow) are measured along the projection of the $\overline{\Gamma \text{K}}$ direction. The out-of-plane spin polarization component ($S_z$, in blue) is measured with both rotator lens modes  (see Experimental Methods and Section IV in the SM for more details). (c) ARPES bandmap along the $\overline{\Gamma \text{K}}$ direction, as extracted from the FS in panel (a). (d) Band dispersion parametrization obtained from equation \ref{eq:2} for E$_{\text{DP}}$ = -340 meV, $\uplambda_{\text{ext}}$ = 10 meV and $\uplambda_{\text{int}}$ = -100 meV. (e) In-plane $\textrm{S}_{x^\prime}$, (f) in-plane $\textrm{S}_{y^\prime}$, and (g) out-of-plane $\textrm{S}_{z}$ spin up and down components (upper panels) and spin polarizations (bottom panels), measured at $\textrm{K}$$_{\text{Gr}}^{\text{off}}$, as indicated in panels (a) and (c). Significant spin splittings for the $\textrm{S}_{x^\prime}$ in-plane component (from 30 meV to 70 meV), and for the $\textrm{S}_{z}$ out-of-plane one (from 30 meV to 40 meV) are observed, while they are negligible for the $\textrm{S}_{y^\prime}$ in-plane component.}
	\label{Fig4}
\end{figure*}

SR-ARPES measurements on the $4/9$ ML Te phase in the vicinity of the DP were carried out with the aim of seeking for any energy spin-splitting or spin polarization of the in-plane or the out-of-plane spin components, in order to gain further insight about the source of the bandgap opening. The mini-Mott spin detector employed in the experiments possesses a spin rotator lens system that allows for the recording of spin-resolved photoemission data for three orthogonal spin components: two spin components parallel to the surface plane ($x^{\prime}$ and $y^{\prime}$ in-plane spin components) and one orthogonal component ($z$ out-of-plane spin component) along the surface normal (see the Experimental Methods, the sketch in Fig.\,\ref{Fig4}(b), and Fig.\,S6 in the SM for the spin components detection geometry and definition).

Measurements were acquired with h$\upnu$ = 21.2 eV and at RT at an emission angle slightly off the $\overline{\textrm{K}}$$_{\text{Gr}}$ point of the Gr SBZ, corresponding to the Fermi wave vector denoted as $\textrm{K}$$_{\text{Gr}}^{\textrm{off}}$, as indicated in the experimental Fermi surface (FS) in Fig.\,\ref{Fig4}(a) (which exhibits the characteristic trigonal symmetry of quasi-freestanding Gr on Ir(111)) and in the ARPES bandmap in Fig.\,\ref{Fig4}(c) by the dashed purple vertical and slanting lines, respectively. Raw spectra (see Fig.\,S6 in the SM) were normalized, and spin up and down components (Figs.\,\ref{Fig4}(e)-(g), upper panels) were obtained by applying equations \ref{Eq_Spin_up} and \ref{Eq_Spin_down} in the Experimental Methods, while the spin polarizations (Figs.\,\ref{Fig4}(e)-(g), bottom panels) were calculated through equation \,\ref{Eq_Spin_pol}. It is important to highlight that the accuracy in the estimation of the energy splittings is not limited by the instrumental resolution, but rather by the acquired statistics, whose associated statistical error, as obtained from the EDC fitting, is of $\sim$\,10 meV.

Both detected in-plane spin components exhibit a sizeable and opposite spin polarization for the $\uppi$ and $\uppi$* bands, as shown in Figs.\,\ref{Fig4}(e) and (f), which are in qualitative agreement with a net chiral in-plane spin component, as illustrated in the sketch in Fig.\,\ref{Fig4}(b). Remarkably, the $x^{\prime}$ in-plane spin component (Fig.\,\ref{Fig4}(e)) shows a significant energy splitting between its up (blue) and down (red) spin components. The values amount for 30 meV and 70 meV for the $\uppi$ and $\uppi$* bands, which are noticeably above the fitting error. For the $z$ out-of-plane spin component, a positive spin polarization and energy spin splitting values of 40 and 30 meV are detected (Fig.\,\ref{Fig4}(g)), which are slightly above the fitting error and whose trends suggest that it constitutes a real effect. 

Thus, a nontrivial chiral spin texture is evidenced for Gr/Te/Ir(111). This non-planar spin texture, which exhibits both in-plane and out-of-plane energy spin splittings, as well as a substantial spin polarization of up to a 20\%, implies the induction of both an extrinsic (Rashba) and an intrinsic (Kane–Mele) SOI in Gr. This finding aligns with the observations made by Otrokov \textit{et al.} \cite{otrokov2018evidence} for Pb intercalated on Gr/Ir(111). However, given that a bandgap has been identified at the DP in both the Te-intercalated phases, the present study suggests that the contribution from Kane and Mele dominates over the Rashba contribution, similar to the case of Pb intercalated on Gr/Pt(111) \cite{klimovskikh2017spin}.

\centering
\subsection*{Discussion on the origin of the bandgap opening} 
\justifying

To validate our interpretation of the SR-ARPES data and the proposed SOI origin for the induced bandgap, the electronic band dispersion has been parametrized following the Hamiltonian describing the relativistic Dirac electronic states of Gr near the $\overline{\text{K}}_{\textrm{Gr}}$ point (H$_{0}$). The Hamiltonian considers both the intrinsic (H$_{\textrm{int}}$, Kane and Mele \cite{kane2005quantum}) and the extrinsic (H$_{\textrm{ext}}$, Rashba \cite{rashba2009graphene}) SOI \cite{abdelouahed2010spin,gmitra2009band}:
\begin{equation} 
\begin{split}
    \textrm{H} = \textrm{H}_{0} + \textrm{H}_{\text{int}} + \textrm{H}_{\text{ext}} =\\
    = \hbar \text{v}_{\text{F}} (\upkappa \upsigma_{\text{x}} \text{k}_{\text{x}} + \upsigma_{\text{y}} \text{k}_{\text{y}}) + \uplambda_{\text{int}} \upkappa \upsigma_{\text{z}} \text{s}_{\text{z}} +\\
    \quad +\uplambda_{\text{ext}} (\upkappa \upsigma_{\text{x}} \text{s}_{\text{y}} - \upsigma_{\text{y}} \text{s}_{\text{x}})
\end{split}
\label{eq:hamiltonian}
\end{equation}

where $\upkappa = \pm 1$ correspond to the cones at $\overline{\text{K}}_{\textrm{Gr}}$ and $\overline{\text{K}}^{\prime}_{\textrm{Gr}}$; $\upsigma_{\text{x,y,z}}$ are the Pauli matrices acting on the pseudospin space formed by the two Gr sublattices; k$_{\text{x,y,z}}$ are the cartesian components of the wave vector of the electron relative to the $\overline{\textrm{K}}_{\textrm{Gr}}$ point; $\uplambda_{\text{ext,int}}$ stand for the strength of the extrinsic (Rashba) and intrinsic (Kane and Mele) SOI; and s$_{\text{x,y,z}}$ are the spin Pauli matrices. Here H$_{\text{int}}$, as compared to H$_{0}$, introduces the intrinsic SOC in the out-of-plane component and lifts the orbital degeneracy at the $\overline{\textrm{K}}_{\textrm{Gr}}$ point, preserving nevertheless the space inversion and the TRS. Accordingly, H$_{\text{ext}}$, as it considers extrinsic SOC, keeps modifying the electronic description of the system and lifts the two-fold degeneracy of the bands, breaking the spatial inversion symmetry. The eigenvalues of the overall Hamiltonian in equation \ref{eq:hamiltonian} near the $\overline{\text{K}}_{\textrm{Gr}}$ point read as:
\begin{equation}
\begin{split}
    \textrm{E}_{\upmu, \upnu} = \textrm{E}_{\textrm{DP}} (\upnu) + \upmu \cdot \uplambda_{\textrm{ext}} +\\
    \quad + \upnu \cdot \sqrt{(\hbar \textrm{v}_{\textrm{F}}\textrm{k})^{2}+(\uplambda_{\textrm{ext}}-\upmu \uplambda_{\textrm{int}})^{2}}
\end{split}
\label{eq:2}
\end{equation}

where $\upmu,\upnu = \pm 1$ correspond to different spin chiralities and to valence and conduction bands, respectively; E$_{\text{DP}} (\upnu)$ to the energy of the DP, which depends on the chiral channel $\upnu$; and the wave vector k is given relative to the $\overline{\text{K}}_{\textrm{Gr}}$ point. In accordance with Kane and Mele's model, a bandgap can be developed if the intrinsic SOI is larger than the Rashba one, i.e., $\lvert\uplambda_{\text{int}}\rvert>\lvert\uplambda_{\text{ext}}\rvert$. In this sense, equation \ref{eq:2} has been parameterized in order to reproduce the experimental band dispersion (Figs.\, \ref{Fig2}(c) and \ref{Fig4}(c)), and the in-plane energy spin splitting around $\overline{\textrm{K}}_{\textrm{Gr}}$. The experimentally obtained value of $\textrm{E}_{\textrm{DP}}$ = -340 meV and SOI strengths of $\uplambda_{\text{ext}}$ = 10 meV and $\uplambda_{\text{int}}$ = -100 meV (of the same order of magnitude as those found by Klimovskikh \textit{et al.} \cite{klimovskikh2017spin}) are found to be the parameters that better reproduce the experimental band dispersion, as they give rise to the model in Fig.\,\ref{Fig4}(d), which agrees with the energy spin splitting observed in the experiments in Fig.\,\ref{Fig4}(e).

This model considers the higher energy spin splitting of $\uppi$* states as compared to $\uppi$ states. The spin splitting resulting from this parametrization for the $\uppi$* states is of 60 meV, which reproduces very well the experimental results (Fig.\,\ref{Fig4}(e)). Note also that the DP energy can depend on the chirality, so introducing an energy shift between opposite chiralities at the DP (i.e., E$_{\textrm{DP}}$$(\upnu$ = -1) $\neq$ E$_{\textrm{DP}}$$(\upnu $ = +1), particularly, E$_{\textrm{DP}}$$(\upnu$ = -1) = -340 meV and E$_{\textrm{DP}}$$(\upnu $ = +1) = -320 meV) would also take into account the spin splitting observed for $\uppi$ states, of 20 meV, without affecting the bandgap amplitude. The model depicted in  Fig.\,\ref{Fig4}(d) is also in agreement with one of the main requirements for the realization of the QSH effect, that is, that the intrinsic SOI is larger than the Rashba one. A similar model was proposed by Klimovskikh \textit{et al.} \cite{klimovskikh2017spin}, where a larger intrinsic SOI was considered, $|\uplambda_{\text{int}}|>|\uplambda_{\text{ext}}|$, resulting in a nontrivial energy gap amounting for $2(|\uplambda_{\text{int}}|-|\uplambda_{\text{ext}}|)$. According to that, the estimated bandgap from the present model would be of $2(|\uplambda_{\text{int}}|-|\uplambda_{\text{ext}}|) = 2 \cdot (100-10) = 180$ meV. Thus, based on this simple band dispersion parametrization, the main experimental findings as the bandgap opening, the nontrivial and non-planar chiral spin texture, and the higher energy splitting for the $\pi^*$ states, are described. 

Gr sublattice symmetry breaking can also originate a bandgap opening, as stated in the Introduction section. In such a case, however, no in-plane spin dependence (neither out-of-plane) should have been detected, as for Bi intercalated on Gr/Ir(111) \cite{krivenkov2021origin}. Furthermore, from the structural analysis extracted out of the LEED and STM (Fig.\,\ref{Fig1}) and, assuming that the Gr lattice constant does not change upon Te intercalation, the superstructures of intercalated Te are commensurate with the underlying Ir(111) and, thus, not with the Gr layer, ruling out the possibility of the sublattice symmetry breaking being at the origin of the bandgap emergence, as in Gr/Bi/Ir(111) \cite{krivenkov2021origin, warmuth2016band}. In this way the sublattice symmetry should be preserved as in the cases of Pb-intercalated systems \cite{otrokov2018evidence,klimovskikh2017spin}. 

Consequently, the intrinsic SOI is proposed as the origin of the bandgap opening at the DP induced by Te intercalation. Thus, the intercalation of Te in Gr/Ir(111) offers a promising platform to tailor SOC properties, including the possibility of generating a bandgap or inducing electronic doping, very powerful tools for engineering, not only Gr systems, but more complex 2D vdW heterostructures as TMDs \cite{gmitra2016trivial,roldan2014momentum,avsar2014spin}.

\centering
\section*{Conclusions}
\justifying

In this work, Te has been effectively intercalated in Gr/Ir(111). Depending on the Te coverage, two different phases have been structurally resolved, both of them commensurate with the underlying Ir(111) substrate. Remarkably, a bandgap opening at the DP of Gr is induced at RT, independent of the structural phase, while the relativistic dispersion of Gr remains undisrupted. The origin of the bandgap has been assigned to a sizeable intrinsic SOI, which exceeds the extrinsic Rashba one. The DP has also been shown to be energetically tunable towards the E$_{\textrm{F}}$ by controlling the amount of intercalated Te, while preserving the bandgap. These results may open a new path for the development of new electronic and spintronic devices, in particular, a QSH insulator phase, whose main requirements have been fulfilled in the investigated heterostructures.

\section*{Acknowledgements}

This work has been supported by the Spanish Ministry of Science and Innovation, Grants no. PID2021-123776NB-C21 (CONPHASETM), PGC2018-098613-B-C21 (SpOrQuMat), EQC2019-006304-P (Equipamiento Científico), PID2020-116181RB-C31 (SOnanoBRAIN), PID2021-128011NB-I00, PID2021-123295NB-I00, and PID2019-105458RB-I00; and by the Comunidad de Madrid through projects S2018/NMT-4511 (NMAT2D) and P2018/NMT-4321 (NANOMAGCOST). IMDEA Nanociencia and IFIMAC acknowledge financial support from the Spanish Ministry of Science and Innovation through ``Severo Ochoa'' (Grant CEX2020-001039-S) and ``María de Maeztu'' (Grant CEX2018-000805-M) Programmes for Centres of Excellence in R$\&$D, respectively. Financial support through the (MAD2D-CM)-MRR MATERIALES AVANZADOS-IMDEA-NC and (MAD2D-CM) MRR MATERALES AVANZADOS-UAM is acknowledged. 

M.G. has received financial support through the ``Ramón y Cajal'' Fellowship program (RYC2020-029317-I) and ``Ayudas para Incentivar la Consolidación Investigadora'' (CNS2022-135175).

\centering
\section*{Experimental Methods}
\justifying

\textbf{Growth and preparation methods:} Samples were prepared onto a Ir (111) monocrystal subjected to several sputtering-annealing cycles prior to the Gr deposition. Gr was grown on Ir(111) by ethylene chemical vapour deposition with a total dose of 108 L. Afterwards a post-annealing at 1443 K during 1 minute was also applied. As a result, a high quality graphene layer of a single rotational phase is obtained as indicated by the LEED and STM images in Fig.\,S1(a) in the SM. Tellurium was evaporated on top of Gr/Ir(111) at RT from a Knudsen cell at T$_{\textrm{Te}}$ = 613 K during betwen 4 and 10 minutes. After that, an annealing at 483 K during 5 minutes was applied in order to promote the intercalation process, confirmed by the recorded LEED patterns in Fig.\,S1(b) in the SM. The characterization by LEED and STM revealed equivalent structures for the low and medium Te coverage samples, whereas different doping levels were observed in the ARPES experiments.

\textbf{STM experiments:} STM measurements were performed in a UHV system equipped with a low temperature STM (LT-STM). STM images were recorded in constant current mode at RT or $77\,$K. The data were processed using the WSxM software \cite{Horcas2007}. 

\textbf{ARPES experiments:} ARPES experiments were carried out at IMDEA Nanociencia. Samples were previously characterized by STM and then transferred to the ARPES chamber by means of a UHV suitcase, being always kept at pressures better than $10^{-8}$ mbar. The analysis chamber was at RT and at a base pressure of $10^{-10}$ mbars.
ARPES measurements were performed with a He lamp with photon energy h$\upnu = 21.2$ eV and a Specs Phoibos 150 hemispherical energy analyzer, with an acceptance angle of $\pm$10º at Medium Angle Mode and $\pm$ 15º at Wide Angle Mode, providing an angular resolution better than 0.05º (0.002 \AA$^{-1}$) and 0.075º (0.003 \AA$^{-1}$), respectively for the short and long energy range maps, and an energy resolution of 80 meV.  

\textbf{SR-ARPES experiments:}
SR-ARPES experiments were carried out at IMDEA Nanociencia. SR-ARPES measurements were recorded with a Combined Out-of-plane Specs 3D Micro-Mott/2D-CCD detector analyzer, which allows for the simultaneous measurement of the three spin components. Two orthogonal in-plane spin components can be measured by an implemented spin rotator lens system, which rotates $\pm$45$^{\circ}$ the in-plane spins: with rotator lens +45$^{\circ}_{(+1)}$ channels 1 and 2 measure one in-plane component (e.g., $x^{\prime}$), and channeltrons 3 and 4 detect the out-of-plane one ($z$); whereas for the -45º$_{(-1)}$ rotator lens it is the remaining in-plane component ($y^{\prime}$) the one which is detected in channels 1 and 2, while channels 3 and 4 redundantly detect the out-of-plane ($z$) signal. 
A circular aperture for the spin-transfer lens of 3 mm, and a pass energy of 10 eV were used, which give rise to energy and angular (momentum) resolutions of 75 meV and 2.25º (0.06 \AA$^{-1}$) \cite{berntsen2010spin}. The spin up and spin down components have been calculated from the results obtained from the normalized and background subtracted spectra in Fig. S6 in the SM by means of equations \ref{Eq_Spin_up} and \ref{Eq_Spin_down}:
\begin{equation}
\textrm{I}_{\uparrow} = ( 1 + \textrm{P}_{\textrm{in(out)}}) \cdot \frac{\textrm{I}_{\textrm{in(out)}}}{2}
\label{Eq_Spin_up}
\end{equation}
\begin{equation}
\ \textrm{I}_{\downarrow} = (1 - \textrm{P}_{\textrm{in(out)}}) \cdot \frac{\textrm{I}_{\textrm{in(out)}}}{2}
\label{Eq_Spin_down}
\end{equation}

where P$_{\textrm{in(out)}}$ is the spin polarization, defined as: 
\begin{equation}
    \textrm{P}_{\textrm{in(out)}} = \frac{\textrm{A}_{\textrm{meas}}}{\textrm{S}_{\textrm{eff}}}
    \label{Eq_Spin_pol}
\end{equation}

being A$_{\textrm{meas}}$ the measured asymmetry, and S$_{\textrm{eff}}$ the Sherman function, which in our experimental setup amounts for 0.16.

\textbf{DFT calculations:}
density functional theory calculations were carried out within the projector augmented wave (PAW) approach \cite{Blochl_PAW}, as implemented in the VASP code \cite{VASP1,VASP2,VASP3}, using the Perdew-Burke-Ernzerhof (PBE) exchange correlation functional \cite{PBE_xc} and the Tkatchenko-Scheffler \cite{TS-vdW} corrections, to account for weak dispersion forces. A 400 eV plane wave cut-off and a total energy threshold of $10^{-5}$ eV for the self consistent field calculations were adopted. Depending on the size of the in-plane unit cell, the BZ sampling was carried out using unshifted Monkhorst-Pack grids \cite{MonkPack}, ensuring a $\Delta \textrm{k} \lesssim$ 0.1 \AA. The Ir surface was modeled by 6-layer-thick Ir(111) slabs. Different amounts of intercalated Te were studied (see the SM for further information) by placing the Te atoms on one side of the slab and a flat and stretched (so that the Gr lattice vectors matched the Ir(111) lattice vectors) graphene layer on top of it.  In all cases a vacuum region of, at least, 20 \AA$ $ in the out of plane direction was adopted. The final geometries were obtained relaxing the coordinates of the Gr atoms, the Te atoms, and the topmost two Ir layers until the maximum force was less than 0.01 eV/\AA. 
The band plots shown in Figs.\,S7 and S8 in the SM were obtained calculating the Kohn-Sham one electron energies and wavefunction over the usual BZ path of Gr. Each electron state is assigned a weight calculated as its overlap with the the wave function of a perfectly flat and accordingly stretched Gr layer. This procedure, in one go, singles out the Gr derived states and allows to get rid of the replicated band due to the band folding, typical of the supercell calculations \cite{BandUnfolding}.

\bibliography{grTeIr.bib}

\end{document}